\documentclass[12pt]{article}
\usepackage{cite}
\usepackage{amssymb} 
\newcommand{\R}{{\rm R}}
\newcommand{\C}{{\rm C}}

\newcommand{\SO}{{\rm SO}}
\newcommand{\SL}{{\rm SL}}
\newcommand{\Sp}{{\rm Sp}}
\newcommand{\ISO}{{\rm ISO}}
\newcommand{\SU}{{\rm SU}}
\newcommand{\U}{{\rm U}}
\newcommand{\Spin}{{\rm Spin}}

\newcommand{\beq}{\begin{equation}}
\newcommand{\eeq}[1]{\label{#1}\end{equation}}
\newcommand{\bea}{\begin{eqnarray}}
\newcommand{\eea}[1]{\label{#1}\end{eqnarray}}

\newcommand{\fft}[2]{{\frac{#1}{#2}}}
\newcommand{\ft}[2]{{\textstyle{\frac{#1}{#2}}}}
\ifx\Bbb\undefined
\else
\renewcommand{\R}{{\mathbb R}}
\renewcommand{\C}{{\mathbb C}}
\fi
\ifx\ltimes\undefined
\newcommand{\ltimes}{{\kern3pt\hbox{\vrule width 0.4pt height 5.30pt depth
.0pt}\kern-1.76pt\times\kern1pt}}
\fi
\begin{document}

\begin{titlepage}
\begin{flushright}
MCTP-03-11\\
hep-th/0303140
\end{flushright}

\vspace{15pt}

\begin{center}
{\large\bf Hidden Spacetime Symmetries and Generalized Holonomy\\
in M-theory%
\footnote{Research supported in part by DOE Grant DE-FG02-95ER40899.}}

\vspace{15pt}

{M.~J.~Duff\footnote{mduff@umich.edu} and
James T.~Liu\footnote{jimliu@umich.edu}}

\vspace{7pt}

{\it Michigan Center for Theoretical Physics\\
Randall Laboratory, Department of Physics, University of Michigan\\
Ann Arbor, MI 48109--1120, USA}

\end{center}

\begin{abstract}
In M-theory vacua with vanishing 4-form $F_{(4)}$, one can invoke the
ordinary Riemannian holonomy $H \subset \SO(10,1)$ to account for
unbroken supersymmetries $n=1$, 2, 3, 4, 6, 8, 16, 32.  However, the
generalized holonomy conjecture, valid for non-zero $F_{(4)}$, can
account for more exotic fractions of supersymmetry, in particular
$16<n<32$.  The conjectured holonomies are given by ${\cal H} \subset
{\cal G}$ where ${\cal G}$ are the generalized structure groups ${\cal
G}=\SO(d-1,1) \times G(spacelike)$, ${\cal G}= \ISO(d-1) \times G(null)$
and ${\cal G}=\SO(d) \times G(timelike)$ with $1\leq d<11$.  For example,
$G(spacelike)=\SO(16)$, $G(null)=[\SU(8) \times \U(1)]\ltimes \R^{56}$
and $G(timelike)=\SO^*(16)$ when $d=3$.  Although extending spacetime
symmetries, there is no conflict with the Coleman-Mandula theorem.
The holonomy conjecture rules out certain vacua which are otherwise
permitted by the supersymmetry algebra.
\end{abstract}

\end{titlepage}


\section{Introduction}
\label{Physics}

M-theory not only provides a non-perturbative unification of the five
consistent superstring theories, but also embraces earlier work on
supermembranes and eleven-dimensional supergravity \cite{World}.  It
is regarded by many as the dreamed-of final theory and has accordingly
received an enormous amount of attention.  It is curious, therefore,
that two of the most basic questions of M-theory have until now remained
unanswered:

i) {\it What are the symmetries of M-theory?}

ii) {\it How many supersymmetries can vacua of M-theory preserve?}

The first purpose of this paper is to argue that M-theory possesses
previously unidentified hidden spacetime (timelike and null) symmetries in
addition to the well-known hidden internal (spacelike) symmetries.  These
take the form of generalized structure groups ${\cal G}$ that replace
the Lorentz group $\SO(10,1)$.

The second purpose is to argue that the number of supersymmetries
preserved by an M-theory vacuum is given by the number of singlets
appearing in the decomposition of the 32-dimensional representation of
${\cal G}$ under ${\cal G} \supset {\cal H}$ where ${\cal H}$ are
generalized holonomy groups.

The equations of M-theory display the
maximum number of supersymmetries $N$=32, and so $n$, the number of
supersymmetries preserved by a particular vacuum, must be some integer
between 0 and 32.  But are some values of $n$ forbidden and, if so,
which ones?  For quite some time it was widely believed that, aside
from the maximal $n=32$, $n$ is restricted to $0\leq n\leq 16$ with
$n=16$ being realized by the fundamental BPS objects of M-theory: the
M2-brane, the M5-brane, the M-wave and the M-monopole.  The subsequent
discovery of intersecting brane configurations with
$n=0$, 1, 2, 3, 4, 5, 6, 8, 16 lent credence to this argument.  In
\cite{Gauntletthull1}, on the other hand, it was shown that all values
$0\leq n \leq 32$ are allowed by the M-theory algebra \cite{Townsend},
and examples of vacua with $16< n < 32$ have indeed since been found.
Following \cite{Duffstelle} and \cite{Duff}, we here
put forward a {\it generalized holonomy conjecture} according to
which the answer lies somewhere in between.  Evidence in favor of
this conjecture includes the observations that there are no known
counterexamples and that a previously undiscovered example predicted in
\cite{Duff}, namely $n$=14, has recently been found \cite{Harmark}.

As we shall see, these conjectures are based on a group-theoretical argument
which applies to the fully-fledged M-theory.  To get the ball rolling,
however, we begin with the low energy limit of M-theory, namely $D=11$
supergravity.  The unique $D=11$ supermultiplet is comprised of
a graviton $g_{MN}$, a gravitino $\Psi_M$ and $3$-form gauge field
$A_{MNP}$, where $M=0, 1, \ldots 10$, with $44$, $128$ and $84$
physical degrees of freedom, respectively. In section \ref{hidden}, we
conjecture that the supergravity equations of motion for this set of
fields admit hidden timelike and null symmetries (in addition to
previously demonstrated hidden spacelike ones).  Then in section
\ref{holonomy} we propose that, so long as the $D=11$ Killing spinor
equation has such hidden symmetries, we may enlarge the tangent space
group into a generalized structure group.  This allows us to analyze
the number of supersymmetries based on a generalized holonomy conjecture.
Partial justification for this conjecture is presented in section
\ref{sec:dimred} in the context of a dimensionally reduced theory.  In
section~\ref{N} we discuss some consequences of generalized holonomy
for classifying supersymmetric vacua, and finally conclude in
section~\ref{M}.


\section{Hidden spacetime symmetries of D=11 supergravity}
\label{hidden}

Long ago, Cremmer and Julia \cite{Cremmerjulia}
pointed out that, when dimensionally reduced to $d$ dimensions, $D=11$
supergravity exhibits hidden symmetries.  For example ${\rm E}_7(global)
\times \SU(8)(local)$ when $d=4$ and ${\rm E}_8(global) \times \SO(16)(local)$
when $d=3$.  The question was then posed \cite{Fradkin}: do these
symmetries appear magically only after dimensional reduction, or were
they already present in the full uncompactified and untruncated $D=11$
theory?  The question was answered by de Wit and Nicolai
\cite{Dewitnicolai2,Nicolai} who made a $d/(11-d)$ split and fixed the
gauge by setting to zero the off-diagonal components of the elfbein.
They showed that in the resulting field equations the local symmetries
are indeed already present, but the global symmetries are not.  For
example, after making the split $\SO(10,1) \supset \SO(3,1) \times
\SO(7)$, we find the enlarged symmetry $\SO(3,1) \times \SU(8)$.  There
is no global ${\rm E}_7$ invariance (although the 70 internal components of
the metric and 3-form may nevertheless be assigned to an ${\rm E}_7/\SU(8)$
coset).  Similar results were found for other values of $d$: in each
case the internal subgroup $\SO(11-d)$ gets enlarged to some compact
group $G(spacelike)$ while the spacetime subgroup $\SO(d-1,1)$ remains
intact%
\footnote{We keep the terminology ``spacetime'' and ``internal''
even though no compactification or dimensional reduction is implied.}.
In this paper we ask instead whether there are hidden {\it spacetime}
symmetries.  This is a question that could have been asked long ago,
but we suspect that people may have been inhibited by the
Coleman-Mandula theorem which forbids combining spacetime and internal
symmetries \cite{Coleman}.  However, this is a statement about
Poincare symmetries of the S-matrix and here we are concerned
with Lorentz symmetries of the equations of motion, so there will be no
conflict.

The explicit demonstration of $G(spacelike)$ invariance by de Wit and
Nicolai is very involved, to say the least.  However, the result is
quite simple: one finds the same $G(spacelike)$ in the full
uncompactified $D=11$ theory as was already found in the spacelike
dimensional reduction of Cremmer and Julia.  Here we content ourselves
with the educated guess that the same logic applies to $G(timelike)$
and $G(null)$: they are the same as what one finds by timelike and
null reduction, respectively. So we propose
that, after making a $d/(11-d)$ split, the Lorentz subgroup
$G=\SO(d-1,1) \times \SO(11-d)$ can be enlarged to the
generalized structure groups ${\cal G}=\SO(d-1,1) \times
G(spacelike)$, ${\cal G}= \ISO(d-1) \times G(null)$ and ${\cal G}= \SO(d)
\times G(timelike)$ as shown in Tables~\ref{sgen}, \ref{ngen}
and \ref{tgen}.

\begin{table}[t]
\begin{center}
\begin{tabular}{c|lc}
$d/(11-d)$&${\cal G}= \SO(d-1,1) \times G(spacelike)$&$\epsilon$ representation\\
\hline
10/1&$\SO(9,1) \times \{1\}$&$16+\overline{16}$\\
9/2&$\SO(8,1) \times \SO(2)$&$16_{\pm1/2}$\\
8/3&$\SO(7,1) \times \SO(3) \times \SO(2)$&$(8_s,2)_{1/2}+(8_c,2)_{-1/2}$\\
7/4&$\SO(6,1) \times \SO(5)$&$(8,4)$\\
6/5&$\SO(5,1) \times \SO(5) \times \SO(5)$&$(4,4,1)+(\overline{4},1,4)$\\
5/6&$\SO(4,1) \times {\rm USp}(8)$&$(4,8)$\\
4/7&$\SO(3,1) \times \SU(8)$&$(2,1,8)+(1,2,\overline{8})$\\
3/8&$\SO(2,1) \times \SO(16)$&$(2,16)$\\
\hline
2/9&$\SO(1,1) \times \SO(16) \times \SO(16)$&$(16,1)_{1/2}+(1,16)_{-1/2}$\\
1/10&$\{1\}   \times \SO(32)$&$32$
\end{tabular}
\end{center}
\caption{Generalized structure groups: spacelike case.  The last column
denotes the representation of $\epsilon$ under ${\cal G}$.}
\label{sgen}
\end{table}

\begin{table}[t]
\begin{center}
\begin{tabular}{c|lc}
$d/(11-d)$&${\cal G}=\ISO(d-1) \times G(null)$&$\epsilon$ representation\\
\hline
10/1&$\ISO(9)$&$16+16$\\
9/2&$\ISO(8) \times \R$&$8_s+8_s+8_c+8_c$\\
8/3&$\ISO(7) \times \ISO(2) \times \R$&$8_{\pm1/2}+8_{\pm1/2}$\\
7/4&$\ISO(6) \times [\SO(3) \times \SO(2)] \ltimes
\R^6_{(3,2)}$&$(4,2)_{\pm_1/2}+(\overline{4},2)_{\pm1/2}$\\
6/5&$\ISO(5) \times \SO(5) \ltimes \R^{10}_{(10)}$&$(4,4)+(4,4)$\\
5/6&$\ISO(4) \times [\SO(5) \times \SO(5)] \ltimes
\R^{16}_{(4,4)}$&$(2,1,4,1)+(2,1,1,4)$\\
&&\qquad$+(1,2,4,1)+(1,2,1,4)$\\
4/7&$\ISO(3) \times {\rm USp}(8)\ltimes \R^{27}_{(27)}$&$(2,8)+(2,8)$\\
3/8&$\ISO(2) \times [\SU(8) \times \U(1)]\ltimes
                        \R^{56}_{(28_{1/2},\overline{28}_{-1/2})}$&$
(8_{1/2})_{\pm1/2}+(\overline{8}_{-1/2})_{\pm1/2}$\\
\hline
2/9&$\R      \times \SO(16) \ltimes \R^{120}_{(120)}$&$16+16$\\
1/10&$\{1\}  \times [\SO(16)\times\SO(16)]\ltimes\R^{256}_{(16,16)}$&$(16,1)
+(1,16)$
\end{tabular}
\end{center}
\caption{Generalized structure groups: null case.  The last column
denotes the representation of $\epsilon$ under the maximum compact
subgroup of ${\cal G}$.}
\label{ngen}
\end{table}

\begin{table}[t]
\begin{center}
\begin{tabular}{c|lc}
$d/(11-d)$&${\cal G}= \SO(d) \times G(timelike)$&$\epsilon$ representation\\
\hline
10/1&$\SO(10) \times \{1\}$&$16+\overline{16}$\\
9/2&$\SO(9) \times \SO(1,1)$&$16_{\pm1/2}$\\
8/3&$\SO(8) \times \SO(2,1) \times \SO(1,1)$&$(8_s,2)_{1/2}+(8_c,2)_{-1/2}$\\
7/4&$\SO(7) \times \SO(3,2)$&$(8,4)$\\
6/5&$\SO(6) \times \SO(5,\C)$&$(4,4)+(\overline{4},\overline{4})$\\
5/6&$\SO(5) \times {\rm USp}(4,4)$&$(4,8)$\\
4/7&$\SO(4) \times \SU^*(8)$&$(2,1,8)+(1,2,\overline{8})$\\
3/8&$\SO(3) \times \SO^*(16)$&$(2,16)$\\
\hline
2/9&$\SO(2) \times \SO(16,\C)$&$16_{1/2}+\overline{16}_{-1/2}$\\
1/10&$\{1\} \times \SO(16,16)$&$32$
\end{tabular}
\end{center}
\caption{Generalized structure groups: timelike case.  The last column
denotes the representation of $\epsilon$ under ${\cal G}$.}
\label{tgen}
\end{table}

Some of the noncompact groups appearing in the Tables may be unfamiliar,
but a nice discussion of their properties may be found in \cite{Gilmore}.
For $d>2$ the groups $G(spacelike)$, $G(timelike)$ and $G(null)$ are
the same as those obtained from the spacelike dimensional reductions
of Cremmer and Julia \cite{Cremmerjulia}, the timelike reductions of
Hull and Julia \cite{Hulljulia}%
\footnote{Actually, for the 8/3 split, we have the factor $\SO(1,1)$
instead of their $\SO(2)$.},
and the null
reduction of section \ref{genhol}, respectively. For our purposes,
however, their physical interpretation is very different.  They are
here proposed as symmetries of the full $D=11$ equations of motion;
there is no compactification involved, whether toroidal or otherwise.
This conjecture that these symmetries are present in the full theory
and not merely in its dimensional reductions may be put to the test,
however, as we shall later describe.  For $d\leq 2$ it is less clear whether
these generalized structure groups are actually hidden symmetries.  See
the caveats of section~\ref{sec:dimred}. The $\SO(16) \times \SO(16)$ for
$d=2$ is also discussed by Nicolai \cite{Nicolai:1991tt}.


\section{Hidden Symmetries and Generalized Holonomy}
\label{holonomy}

We begin by reviewing the connection between holonomy and the number of
preserved supersymmetries, $n$, of supergravity vacua.  This also serves
to define our notation.  Subsequently, we introduce a generalized holonomy
which involves the hidden symmetries conjectured in the previous
section.

\subsection{Riemannian Holonomy}

We are interested in solutions of the bosonic field equations
\begin{equation}
R_{MN}=\frac{1}{12}\left(F_{MPQR}F_{N}{}^{PQR}-\frac{1}{12}g_{MN}
F^{PQRS}F_{PQRS}\right)
\end{equation}
and
\begin{equation}
d*\!F_{(4)}+\fft12F_{(4)} \wedge F_{(4)}=0,
\end{equation}
where $F_{(4)}=dA_{(3)}$.  The supersymmetry transformation rule of the
gravitino reduces in a purely bosonic background to
\begin{equation}
\delta \Psi_{M}={\tilde D}_{M} \epsilon,
\label{eq:11gto}
\end{equation}
where the parameter $\epsilon$ is a 32-component anticommuting spinor,
and where
\begin{equation}
\label{covariant}
{\tilde D}_{M}=D_{M}-
\frac{1}{288}(\Gamma_M{}^{NPQR}-8\delta_M^N\Gamma^{PQR})F_{NPQR},
\label{supercovariant}
\end{equation}
where $\Gamma^{A}$ are the $D=11$ Dirac matrices.  Here $D_{M}$ is the
usual Riemannian covariant derivative involving the connection $\omega_{M}$
of the usual structure group $\Spin(10,1)$, the double cover of $\SO(10,1)$,
\begin{equation}
D_{M}=\partial_{M}+\frac{1}{4}\omega_{M}{}^{AB}\Gamma_{AB}.
\end{equation}
The number of supersymmetries preserved by an M-theory background depends
on the number of covariantly constant spinors,
\begin{equation}
{\tilde D}_{M}\epsilon=0,
\end{equation}
called {\it Killing} spinors.  It is the presence of the terms involving
the 4-form $F_{(4)}$ in (\ref{covariant}) that makes this counting difficult.
So let us first examine the simpler vacua for which $F_{(4)}$ vanishes.
Killing spinors then satisfy the integrability condition
\begin{equation}
[{D}_{M}, {D}_{N}] \epsilon=\frac{1}{4}R_{MN}{}^{AB}\Gamma_{AB}\epsilon=0,
\label{integrability}
\end{equation}
where $R_{MN}{}^{AB}$ is the Riemann tensor.  The subgroup of
$\Spin(10,1)$ generated by this linear combination of $\Spin(10,1)$ generators
$\Gamma_{AB}$ corresponds to the ${\it holonomy}$ group ${H}$ of the
connection $\omega_{M}$.  The number of supersymmetries, $n$, is then
given by the number of singlets appearing in the decomposition of the
$32$ of $\Spin(10,1)$ under ${H}$.  In Euclidean signature, connections
satisfying (\ref{integrability}) are automatically Ricci-flat and
hence solve field equations when $F_{(4)}=0$.  In Lorentzian signature,
however, they need only be Ricci-null \cite{Fig} so Ricci-flatness has
to be imposed as an extra condition.  In Euclidean signature, the
holonomy groups have been classified \cite{Berger}.  In Lorentzian
signature, much less is known but the question of which subgroups ${H}$
of $\Spin(10,1)$ leave a spinor invariant has been answered in
\cite{Bryant}.  There are two sequences according as the vector
$v_{A}=\overline{\epsilon}\,\Gamma_{A}\epsilon$ is timelike or null, as shown
in Tables~\ref{static} and \ref{wave}. Since $v^{2} \leq 0$, the
spacelike $v_A$ case does not arise. The timelike $v_A$ case corresponds
to static vacua, where ${H} \subset \Spin(10) \subset \Spin(10,1)$ while
the null case to non-static vacua where ${H} \subset \ISO(9) \subset
\Spin(10,1)$.  It is then possible to determine the possible $n$-values
\cite{Acharya:1998yv,Acharya:1998st} and one finds $n=2$, 4, 6, 8,
16, 32 for static vacua, as shown in Table~\ref{static}, and $n=1$,
2, 3, 4, 8, 16, 32 for non-static vacua, as shown in
Table~\ref{wave}.

\begin{table}[t]
\begin{center}
\begin{tabular}{c|cc}
$d/(11-d)$&$H \subset \SO(11-d)\subset\Spin(10)$& $n$\\
\hline
7/4&$\SU(2) \cong \Sp(2)$& $16$\\
5/6&$\SU(3)$& $8$\\
4/7&${\rm G}_2$& $4$\\
3/8&$\SU(2) \times \SU(2)$& $8$\\
&$\Sp(4)$& $6$\\
&$\SU(4)$& $4$\\
&$\Spin(7)$& $2$\\
1/10&$\SU(2) \times \SU(3)$& $4$\\
&$\SU(5)$& $2$
\end{tabular}
\end{center}
\caption{Holonomy of static M-theory vacua with $F_{(4)}=0$ and their
supersymmetries.}
\label{static}
\end{table}

\begin{table}[t]
\begin{center}
\begin{tabular}{c|cc}
$d/(11-d)$&$H \subset \ISO(d-1)\times\ISO(10-d)\subset\Spin(10,1)$& $n$\\
\hline
10/1&$\R^9$ & $16$\\
6/5&$\R^5\times(\SU(2) \ltimes \R^4)$ & $8$\\
4/7&$\R^3\times(\SU(3) \ltimes \R^6)$ & $4$\\
3/8&$\R^2\times({\rm G}_2 \ltimes \R^7)$ & $2$\\
2/9&$\R\times(\SU(2) \ltimes \R^4) \times (\SU(2) \ltimes \R^4)$ & $4$\\
&$\R\times(\Sp(4) \ltimes \R^8)$ & $3$\\
&$\R\times(\SU(4) \ltimes \R^8)$ & $2$\\
&$\R\times(\Spin(7) \ltimes \R^8)$ & $1$
\end{tabular}
\end{center}
\caption{Holonomy of non-static M-theory vacua with $F_{(4)}=0$ and their
supersymmetries.}
\label{wave}
\end{table}

\subsection{Generalized holonomy}
\label{genhol}

When we want to include vacua with $F_{(4)}\neq 0$ we face the problem
that the connection in (\ref{covariant}) is no longer the spin
connection to which the bulk of the mathematical literature on
holonomy groups is devoted.  In addition to the $\Spin (10,1)$
generators $\Gamma_{AB}$, it is apparent from (\ref{supercovariant})
that there are terms involving $\Gamma_{ABC}$ and $\Gamma_{ABCDE}$.
As a result, the connection takes its values in the full $D=11$
Clifford algebra.  Moreover, this connection can preserve exotic
fractions of supersymmetry forbidden by the Riemannian connection.
For example, the M-branes at angles in \cite{Ohta} include $n$=5, the
11-dimensional pp-waves in
\cite{Michelson,Cvetic:2002si,Gauntletthull2,Bena} include $n=18$, 20,
22, 24, 26 (and $n=28$ for Type IIB), the squashed $N(1,1)$ spaces in
\cite{Page} and the M5-branes in a pp-wave background in \cite{Singh}
include $n$=12 and the G\"{o}del universes in \cite{Gauntlett:2002nw}
include $n=18$, 20, 22, 24.

However, we can attempt to quantify this in terms of generalized
holonomy groups ${\cal H} \subset {\cal G}$ where ${\cal G}$ are
the generalized structure groups discussed in section \ref{hidden}.
The generalized holonomy conjecture
\cite{Duffstelle,Duff} states that one can assign a holonomy ${\cal H}
\subset {\cal G}$ to the generalized connection%
\footnote{A related conjecture was made in \cite{Berkooz}, where the
generalized holonomy could be any subgroup of $\SO(16,16)$.  This also
appears in our conjectured hidden structure groups under the 1/10 split,
though only in the timelike case ${\cal G}(timelike)$.}
appearing in the supercovariant
derivative (\ref{covariant}). Here we propose that, after making
a $d/(11-d)$ split, the Lorentz subgroup $G=\SO(d-1,1) \times \SO(11-d)$
can be enlarged to the generalized structure groups ${\cal
G}=\SO(d-1,1) \times G(spacelike)$, ${\cal G}= \ISO(d-1) \times
G(null)$ and ${\cal G}= \SO(d) \times G(timelike)$ as shown in
Tables~\ref{sgen}, \ref{ngen} and \ref{tgen}.
Note that in the right hand column of the tables we have listed the
corresponding ${\cal G}$ representations under which the 32 supersymmetry
parameters $\epsilon$ transform.  The number of supersymmetries preserved
by an M-theory vacuum is then given by the number of singlets
appearing in the decomposition of these representations under ${\cal G}
\supset {\cal H}$.


\section{Structure groups from dimensional reduction}
\label{sec:dimred}

In this section we provide partial justification for the conjectured
hidden symmetries by demonstrating their presence in the gravitino variation
of the dimensionally
reduced theory.  In particular, we consider a spacelike dimensional
reduction corresponding to a $d/(11-d)$ split.  Turning on only
$d$-dimensional scalars, the reduction ansatz is particularly simple
\begin{equation}
g^{(11)}_{MN}=\pmatrix{\Delta^{-1/(d-2)}g_{\mu\nu}&0\cr0&g_{ij}},\qquad
A^{(11)}_{ijk}=\phi_{ijk},
\end{equation}
where $\Delta=\det{g_{ij}}$.  For $d\le5$, we must also consider the
possibility dualizing either $F_{(4)}$ components or (for $d=3$)
Kaluza-Klein vectors to scalars.  We will return to such possibilities
below.  But for now we focus on $d\ge6$.  In this case, a standard
dimensional reduction of the $D=11$ gravitino transformation,
(\ref{eq:11gto}), yields the $d$-dimensional gravitino transformation
\begin{equation}
\delta\psi_\mu=[D_\mu+\ft14Q_\mu{}^{ab}\Gamma_{ab}+\ft1{24}\partial_\mu
\phi_{ijk}\Gamma^{ijk}]\epsilon.
\label{eq:dgto}
\end{equation}
For completeness, we also note that the $d$-dimensional dilatinos
transform according to
\begin{equation}
\delta\lambda_i=-\ft12\gamma^\mu[P_{\mu\,ij}\Gamma^j
-\ft1{36}(\Gamma_i{}^{jkl}-6\delta_i^j\Gamma^{kl})\partial_\mu\phi_{jkl}]
\epsilon.
\end{equation}
In the above, the lower dimensional quantities are related to their $D=11$
counterparts through
\begin{eqnarray}
&&\psi_\mu=\Delta^{\fft1{4(d-2)}}\left(\Psi^{(11)}_\mu+\fft1{d-2}\gamma_\mu
\Gamma^i\Psi^{(11)}_i\right),\qquad
\lambda_i=\Delta^{\fft1{4(d-2)}}\Psi^{(11)}_i,\nonumber\\
&&\epsilon=\Delta^{\fft1{4(d-2)}}\epsilon^{(11)},\nonumber\\
&&Q^{ab}_\mu=e^{i[a}\partial_\mu e_i{}^{b]},\qquad
P_{\mu\,ij}=e_{(i}^a\partial_\mu e_{j)\,a}.
\end{eqnarray}

We now see that the lower dimensional gravitino transformation,
(\ref{eq:dgto}), may be written in terms of a covariant derivative
under a generalized connection
\begin{equation}
\delta\psi_\mu=\hat D_\mu\epsilon,\qquad
\hat D_\mu=\partial_\mu+\ft14\Omega_\mu,
\end{equation}
where
\begin{equation}
\Omega_\mu=\omega_\mu{}^{\alpha\beta}\gamma_{\alpha\beta}
+Q_\mu{}^{ab}\Gamma_{ab}+\ft1{3!}e^{ia}e^{jb}e^{kc}\partial_\mu\phi_{ijk}
\Gamma_{abc}.
\label{eq:gencon}
\end{equation}
Here $\gamma_\alpha$ are $\SO(d-1,1)$ Dirac matrices, while $\Gamma_a$
are $\SO(11-d)$ Dirac matrices.  This decomposition is suggestive of a
generalized structure group with connection given by $\Omega_\mu$.
However one additional requirement is necessary before declaring this an
enlargement of $\SO(d-1,1)\times \SO(11-d)$, and that is to ensure that the
algebra generated by $\Gamma_{ab}$ and $\Gamma_{abc}$ closes within
itself.  Along this line, we note that the commutators of these internal
Dirac matrices have the schematic structure
\begin{equation}
[\Gamma^{(2)},\Gamma^{(2)}]=\Gamma^{(2)},\qquad
[\Gamma^{(2)},\Gamma^{(3)}]=\Gamma^{(3)},\qquad
[\Gamma^{(3)},\Gamma^{(3)}]=\Gamma^{(6)}+\Gamma^{(2)}.
\label{eq:diralg}
\end{equation}
Here the notation $\Gamma^{(n)}$ indicates the antisymmetric product of
$n$ Dirac matrices, and the right hand sides of the commutators only
indicate what possible terms may show up.  The first commutator above
merely indicates that the $\Gamma_{ab}$ matrices provide a
representation of the Riemannian $\SO(11-d)$ structure group.

For $d\ge6$, the internal space is restricted to five or fewer
dimensions.  In this case, the antisymmetric product $\Gamma^{(6)}$
cannot show up, and the algebra clearly closes on $\Gamma^{(2)}$ and
$\Gamma^{(3)}$.  Working out the extended structure groups for these
cases results in the expected Cremmer and Julia groups listed in the
first four lines of Table~\ref{sgen}.  A similar analysis follows for
$d\le5$.  However, in this case, we must also dualize an additional set
of fields to see the hidden symmetries.  For $d=5$, an additional scalar
arises from the dual of $F_{\mu\nu\rho\sigma}$; this yields an addition
to (\ref{eq:gencon}) of the form $\Omega_\mu^{\rm additional}=\fft1{4!}
\epsilon_\mu{}^{\nu\rho\sigma\lambda} F_{\mu\nu\rho\sigma}\Gamma_{123456}$.
This $\Gamma^{(6)}$ term is precisely what is necessary for the closure
of the algebra of (\ref{eq:diralg}).  Of course, in this case, we must
also make note of the additional commutators
\begin{equation}
[\Gamma^{(2)},\Gamma^{(6)}]=\Gamma^{(6)},\qquad
[\Gamma^{(3)},\Gamma^{(6)}]=\Gamma^{(7)}+\Gamma^{(3)},\qquad
[\Gamma^{(6)},\Gamma^{(6)}]=\Gamma^{(10)}+\Gamma^{(6)}+\Gamma^{(2)}.
\label{eq:gam6com}
\end{equation}
However neither $\Gamma^{(7)}$ nor $\Gamma^{(10)}$ may show up in $d=5$ for
dimensional reasons.

The analysis for $d=4$ is similar; however here
$\Omega_\mu^{\rm
additional}=\fft1{3!}\epsilon_\mu{}^{\nu\rho\sigma}e^{ia}F_{\nu\rho\sigma i}
\Gamma_a\Gamma_{1234567}$.  Closure of the algebra on $\Gamma^{(2)}$,
$\Gamma^{(3)}$ and $\Gamma^{(6)}$ then follows because, while $\Gamma^{(7)}$
may in principle arise in the middle commutator of (\ref{eq:gam6com}),
it turns out to be kinematically forbidden.  For $d=3$, on the other
hand, in additional to a contribution $\Omega_\mu^{\rm additional}
=\fft1{2!\cdot2!}\epsilon_\mu{}^{\nu\rho}e^{ia}e^{jb}F_{\nu\rho ij}
\Gamma_{ab}\Gamma_{12345678}$, one must also dualize the Kaluza-Klein
vectors $g_\mu{}^i$.  Doing so gives rise to a $\Gamma^{(7)}$ in the
generalized connection which, in addition to the previously identified
terms, completes the internal structure group to $\SO(16)$.

The remaining two cases, namely $d=2$ and $d=1$, fall somewhat outside the
framework presented above.  This is because in these low dimensions the
generalized connections $\Omega_\mu$ derived via reduction are partially
incomplete.  For $d=2$, we find
\begin{equation}
\Omega_\mu^{(d=2)}=\omega_\mu{}^{\alpha\beta}
\gamma_{\alpha\beta}+Q_\mu{}^{ab}\Gamma_{ab}+\ft19(\delta_\mu^\nu-
\ft12\gamma_\mu{}^\nu)e^{ia}e^{jb}e^{kc}\partial_\nu\phi_{ijk}\Gamma_{abc},
\label{eq:d2omega}
\end{equation}
where $\gamma_{\mu\nu}=-\fft12\epsilon_{\mu\nu}
(\epsilon^{\alpha\beta}\gamma_{\alpha\beta})$ is necessarily proportional
to the two-dimensional chirality matrix.  Hence from a two-dimensional
point of view, the scalars from the metric enter non-chirally, while the
scalars from $F_{(4)}$ enter chirally.  Taken together, the generalized
connection (\ref{eq:d2omega}) takes values in $\SO(16)_+\times
\SO(16)_-$, which we regard as the enlarged structure group.  However not
all generators are present because of lack of chirality in
the term proportional to $Q_\mu{}^{ab}$.  Thus at this point the
generalized structure group deviates from the hidden symmetry group,
which would be an infinite dimensional subgroup of affine ${\rm E}_8$.
Similarly, for $d=1$, closure of the connection $\Omega_\mu^{(d=1)}$
results in an enlarged $\SO(32)$ structure group.  However this is not
obviously related to any actual hidden symmetry of the $1/10$ split.

Until now, we have considered the spacelike reductions leading to the
generalized structure groups of Table~\ref{sgen}.  For a timelike
reduction, we simply interchange a time and a space direction in the
above analysis%
\footnote{By postulating that the generalized structure groups survive as
hidden symmetries of the full uncompactified theory, we avoid the undesirable
features associated with compactifications including a timelike direction
such as closed timelike curves.}.
This results in an internal Clifford algebra with signature $(10-d,1)$,
and yields the extended symmetry groups indicated in Table~\ref{tgen}.
Turning finally to the null case, we may replace one of the internal
Dirac matrices with $\Gamma_+$ (where $+$, $-$ denote light-cone
directions).  Since $(\Gamma_+)^2=0$, this indicates that the extended
structure groups for the null case are contractions the corresponding
spacelike (or timelike) groups.  In addition, by removing $\Gamma_+$
from the set of Dirac matrices, we essentially end up in the case of one
fewer compactified dimensions.  As a result, the $G(null)$ group in
$d$-dimensions must have a semi-direct product structure involving the
$G(spacelike)$ group in $(d+1)$-dimensions.  Of course, these groups
also contain the original $\ISO(10-d)$ structure group as a subgroup.
The resulting generalized structure groups are given in
Table~\ref{ngen}%
\footnote{The reduction of $D$-dimensional pure gravity along a single
null direction was analyzed by Julia and Nicolai \cite{Julianicolai}.}.
%


\section{Counting supersymmetries}
\label{N}

Having defined a generalized holonomy for vacua with $F_{(4)}\ne0$, we
now turn to some elementary examples.  For the basic objects of
M-theory, the M2-brane configuration may be placed under the 3/8 (spacelike)
classification, as it has three longitudinal and eight transverse directions.
Focusing on the transverse directions (which is the analog of looking at
$\hat D_\mu$), the M2-brane has generalized holonomy $\SO(8)$ contained in
$\SO(2,1)\times\SO(16)$ \cite{Duffstelle}.  In this case, the spinor
decomposes as
$(2,16)=2(8)+16(1)$, indicating the expected presence of 16 singlets.
For the M5-brane with 6/5 (spacelike) split, the generalized $\hat D_\mu$
holonomy is
given by $\SO(5)_+\subset \SO(5,1)\times\SO(5)_+\times\SO(5)_-$, with
the spinor decomposition $(4,4,1)+(\overline{4},1,4)=4(4)+16(1)$.  Since
the wave solution depends on nine space-like coordinates, we may regard
it as a 1/10 (null) split.  In this case, it has generalized $\tilde D_M$
holonomy
$\R^{9}\subset [\SO(16)\times\SO(16)]\ltimes \R^{256}_{(16,16)}$.  The
spinor again decomposes into 16 singlets.  Note, however, that since
the wave is pure geometry, it could equally well be categorized under a
10/1 split as $\R^9\subset \ISO(9)$.  Finally, the KK monopole
is described by a 7/4 (spacelike) split, and has $\hat D_\mu$ holonomy
$\SU(2)_+ \subset \SO(6,1)\times\SO(5)$, where the spinor decomposes as
$(8,4)=8(2)+16(1)$.  In all four cases, the individual objects
preserve exactly half of the 32 supersymmetries.  However each object
is associated with its own unique generalized holonomy, namely
$\SO(8)$, $\SO(5)$, $\R^{9}$ and $\SU(2)$ for the M2, M5, MW and MK,
respectively.

The supersymmetry of intersecting brane configurations may be understood
in a similar manner based on generalized holonomy.  For example, for a
M5 and MK configuration sharing six longitudinal directions, we may
choose a 6/5 split.  In this case, the structure group is
$\SO(5,1)\times \SO(5)_+\times \SO(5)_-$, and the $\hat D_\mu$ holonomies
of the
individual objects are $\SO(5)_+$ and $\SU(2)\subset\SO(5)_{\rm diag}$,
respectively.  The holonomy for the combined configuration turns out to
be $\SO(5)_+\times\SU(2)_-$, with the spinor decomposing as
$(4,4,1)+(\overline{4},1,4)=4(4,1)+4(1,2)+8(1,1)$.  The resulting eight
singlets then signify the presence of a $1/4$ supersymmetric configuration.
In principle, this analysis may be applied to more general brane
configurations.  However one goal of understanding enlarged holonomy is
to obtain a classification of allowed holonomy groups and, as a result, to
obtain a unified treatment of counting supersymmetries.  We now provide
some observations along this direction.

We first note the elementary fact that a $p$-dimensional representation
can decompose into any number of singlets between 0 and $p$, {\it
except} $(p-1)$, since if we have $(p-1)$ singlets, we must have $p$.
It follows that in theories with $N$ supersymmetries, $n=N-1$
is ruled out, even though it is permitted by the supersymmetry algebra.

In some cases, additional restrictions on $n$ may be obtained.  For example,
if the supersymmetry charge transforms as the $(2,16)$ representation of
${\cal G}$ when $d=3$,
then $n$ is restricted to 0, 2, 4, 6, 8, 10, 12, 14, 16, 18, 20, 22, 24,
26, 28, 32 as first noted in \cite{Duff}.  No new values of $n$ are generated
by $d>3$ reps.  For example, the 4 of $\SO(5)$ can decompose only into
0, 2 or 4 singlets but not 1.

We note that all the even values of $n$ discussed so far appear in the
list and that $n=30$ is absent.  This is consistent with the presence of
pp-waves with $n=16$, 18, 20, 22, 24, 26 (and $n=28$ for Type IIB) but the
absence of $n$=30 noted in
\cite{Bena,Michelson,Cvetic:2002si,Gauntletthull2}.  Of course a good
conjecture should not only account for the existing data but should go
on to predict something new.  For example, Gell-Mann's flavor $\SU(3)$
not only accounted for the nine known members of the baryon decuplet
but went on to predict the existence of the $\Omega^{-}$, which was
subsequently discovered experimentally.  For M-theory supersymmetries,
the role of the $\Omega^{-}$ is played by $n=14$ which at the time of
its prediction had not been discovered ``experimentally''.  We note
with satisfaction, therefore, that this missing member has recently been
found in the form of a G\"{o}del universe \cite{Harmark}.

The $d=2$ and $d=1$ cases are more problematic since $\SO(16)\times\SO(16)$
and $\SO(32)$ in principle allow any $n$ except $n=31$.  So more work is
required to explain the presence of M-branes at angles with $n=0$, 1,
2, 3, 4, 5, 6, 8, 16 but the absence of $n=7$ noted in \cite{Ohta}.
Presumably, a more detailed analysis will show that only those
subgroups compatible with these allowed values of $n$ actually appear as
generalized holonomy groups.  The beginnings of a classification of
all supersymmetric $D=11$ solutions may be found in \cite{Gauntlett:2002fz}.

We can apply similar logic to theories with fewer than 32
supersymmetries.  Of course, if M-theory really underlies all
supersymmetric theories then the corresponding vacua will all be
special cases of the above.  However, it is sometimes useful to focus
on such a sub-theory, for example the Type I and heterotic strings
with $N=16$.  Here $G(spacelike)= \SO(d) \times \SO(d)$, $G(null)=
\ISO(d-1) \times \ISO(d-1)$ and $G(timelike)=\SO(d-1,1) \times
\SO(d-1,1)$.  If the supersymmetry charge transforms as a $(2,8)$
representation of the generalized structure group when $d=3$, then $n$
is restricted to 0, 2, 4, 6, 8, 10, 12, 16.  No new values of $n$ are
generated from other $d>4$ reps. Once again, the $d=2$ and $d=1$ cases
require a more detailed analysis.


\section{The full M-theory}
\label{M}

We have focused on the low energy limit of M-theory, but
since the reasoning that led to the conjecture is based just on group
theory, it seems reasonable to promote it to the full
M-theory%
\footnote{Similar conjectures can be applied to M-theory in signatures
(9,2) and (6,5) \cite{Blencowe:1988sk}, the so-called M$^\prime$ and M$^*$
theories \cite{Hull:1998ym}, but the groups will be different.}.
When counting the $n$ value of a particular vacuum, however, we should be
careful to note the phenomenon of {\it supersymmetry without
supersymmetry}, where the supergravity approximation may fail to
capture the full supersymmetry of an M-theory vacuum. For example,
vacua related by T-duality and S-duality must, by definition, have the
same $n$ values. Yet they can appear to be different in
supergravity \cite{DLP1,DLP2}, if one fails to take into account winding
modes and non-perturbative solitons. So more work is needed to verify
that the $n$ values found so far in $D=11$ supergravity exhaust those of
M-theory, and to prove or disprove the conjecture.


\section*{Notes added}

After this paper was posted on the archive, a very interesting paper by
Hull appeared \cite{Hull:2003mf} which generalizes and extends the
present theme.  Hull conjectures that the hidden symmetry of M-theory is
as large as $\SL(32,\R)$ and that this is necessary in order to accommodate
all possible generalized holonomy groups.  We here make some remarks in
the light of Hull's paper:

\newpage
{\it Hidden symmetries}:

Hull stresses that, as a candidate hidden symmetry, $\SL(32,\R)$
is background independent.  However, the hidden symmetries displayed in
Tables~\ref{sgen}, \ref{ngen} and \ref{tgen} are also background
independent.  They depend only on the choice of non-covariant split and gauge
in which to write the field equations.  Hull's proposal is nevertheless
very attractive since $\SL(32,\R)$ contains all the groups in
Tables~\ref{sgen}, \ref{ngen} and \ref{tgen} as subgroups and would
thus answer the question of whether all these symmetries are present
at the same time.

One can accommodate
$\SL(32,\R)$ by extending the $d/(11-d)$ split to include the $d=0$
case.  Then the same $\SL(32,\R)$ would appear in all three tables.  At
the other end, one could also include the $d=11$ case.  Then the same
$\SO(10,1)$ would appear in all three tables.  Our reason for not
including the $d=0$ case stems from the apparent need to make a
non-covariant split and to make the corresponding gauge choice before
the hidden symmetries become apparent \cite{Dewitnicolai2,Nicolai}.
Moreover, from the point of view of guessing the hidden symmetries
from the dimensional reduction, the $d=0$ case would be subject to the
same caveats as the $d=1$ and $d=2$ cases: not all group generators
are present in the covariant derivative.  $\SL(32,\R)$ requires
$\{\Gamma^{(1)},\Gamma^{(2)},
\Gamma^{(3)},\Gamma^{(4)},\Gamma^{(5)}\}$ whereas only
$\{\Gamma^{(2)}, \Gamma^{(3)},\Gamma^{(5)}\}$ appear in the covariant
derivative.  This is an important issue deserving of further study.
That M-theory could involve a ${\rm GL}(32,\R)$ has also been conjectured by
Barwald and West \cite{Barwald:1999is}.

{\it Generalized holonomy}:

Hull goes on to stress the importance of $\SL(32,\R)$ by finding solutions
whose holonomy is contained in $\SL(32,\R)$ but not in Tables~\ref{sgen},
\ref{ngen} and \ref{tgen}. Although not all generators are present in the
covariant derivative, they are all present in the commutator. So we agree
with Hull that $\SL(32,\R)$ is necessary if one wants to embrace all
possible generalized holonomies.

Indeed, since the basic objects of M-theory discussed in section \ref{N}
involve warping by a harmonic function, the $\hat D$ holonomy is smaller
than the $\tilde D$ holonomy, which requires extra $\R^n$ factors.
Interestingly enough, the $\hat D$ holonomy nevertheless yields the
correct counting of supersymmetries.

Hull points out that, in contrast to the groups appearing in Tables~\ref{sgen},
\ref{ngen} and \ref{tgen}, $\SL(32,\R)$ does not obey the $n \neq N-1$
rule of section \ref{N}, and hence M-theory vacua with $n=31$ are in
principle possible%
\footnote{The case for $n=31$ has also been made by Bandos {\it et al.}
\cite{Bandos:2001pu} in the different context of hypothetical preons of
M-theory preserving 31 out of 32 supersymmetries.}.
Of course we do not yet know whether the required $\R^{31}$
holonomy actually appears.  To settle the issue of which $n$ values
are allowed, it would be valuable to do for supergravity what Berger
\cite{Berger} did for gravity and have a complete classification of
all possible generalized holonomy groups. But this may prove quite
difficult.

So we remain open-minded about a formulation of M-theory with
$\SL(32,\R)$ symmetry, but acknowledge the need for $\SL(32,\R)$ from
the point of view of generalized holonomy.


\section*{Acknowledgments}

We have enjoyed useful conversations with Hisham Sati.


\end{document}